# To Use or to Refuse? Re-Centering Student Agency with Generative AI in Engineering Design Education

Thijs Willems
*Lee Kuan Yew Centre for Innovative Cities*
*Singapore University of Technology and Design*
Singapore, Singapore
thijs_willems@sutd.edu.sg

Sumbul Khan
*Science, Mathematics and Technology Singapore University of Technology and Design*
Singapore, Singapore
sumbul_khan@sutd.edu.sg

Qian Huang
*Lee Kuan Yew Centre for Innovative Cities*
*Singapore University of Technology and Design*
Singapore, Singapore
qian_huang@sutd.edu.sg

Bradley Camburn
*Engineering Product Development*
*Singapore University of Technology and Design*
Singapore, Singapore
bradley_camburn@sutd.edu.sg

Nachamma Sockalingam
*Office of Strategic Planning*
*Singapore University of Technology and Design*
Singapore, Singapore
nachamma@sutd.edu.sg

King Wang Poon
*Lee Kuan Yew Centre for Innovative Cities*
*Singapore University of Technology and Design*
Singapore, Singapore
poonkingwang@sutd.edu.sg

*Abstract*— This pilot study traces students' reflections on the use of AI in a 13-week foundational design course enrolling over 500 first-year engineering and architecture students at the Singapore University of Technology and Design. The course was an AI-enhanced design course, with several interventions to equip students with AI based design skills, including (1) AI-based design methods were introduced for the entire design process, (2) An 'AI x Design thinking' competition was incorporated, to encourage students to reflect on their AI use. Students were required to reflect on whether the technology was used as a tool (instrumental assistant), a teammate (collaborative partner), or neither (deliberate non-use). By foregrounding this three-way lens, students learned to use AI for innovation rather than just automation and to reflect on agency, ethics, and context rather than on prompt crafting alone.

Evidence stems from coursework artefacts: thirteen structured reflection spreadsheets and eight illustrated briefs submitted, combined with notes of teachers and researchers. Qualitative coding of these materials reveals shared practices brought about through the inclusion of Gen-AI, including accelerated prototyping, rapid skill acquisition, iterative prompt refinement, purposeful "switch-offs" during user research, and emergent routines for recognizing hallucinations.

Unexpectedly, students not only harnessed Gen-AI for speed but (enabled by the tool–teammate–neither triage) also learned to reject its outputs, invent their own hallucination fire-drills, and divert the reclaimed hours into deeper user research, thereby transforming efficiency into innovation. The implications of the approach we explore shows that: we can transform AI uptake into an assessable design habit; that rewarding selective non-use cultivates hallucination-aware workflows; and, practically, that a coordinated bundle of tool access, reflection, role tagging, and public recognition through competition awards allows AI based innovation in education to scale without compromising accountability.

## I. Introduction

The release of ChatGPT in November 2022 marked a turning point in the integration of generative artificial intelligence (Gen-AI) into higher education [1]. Within a matter of months, students across disciplines began experimenting with large language models to draft essays, summarize readings, debug code, and explore novel ideas [2, 3]. These tools, once viewed as speculative or niche, rapidly became ubiquitous on university campuses. In engineering programs in particular, where structured problem-solving, prototyping, and rapid iteration are central to the learning process, Gen-AI emerged not only as a productivity enhancer but also as a pedagogical disruptor, indicating both challenges [4] and opportunities [5].

Institutional responses were often ambivalent, shaped by concerns over academic integrity, skill erosion, and the difficulty of assessing AI-mediated work [6, 7]. While these concerns are valid, they risk narrowing the scope of educational inquiry to compliance and containment. With its prime focus on whether students use Gen-AI and a more normative stance on whether students should use it in their learning, what remains underexplored in these studies is *how* students actually position it within their learning practice, and what it means to teach this positioning as a skill.

Much of the emerging discourse has drawn on an automation-versus-augmentation framework [8]. In this view, Gen-AI either automates repetitive or mechanical tasks to free up cognitive resources, or augments complex problem-solving through support and suggestion [9, 10]. While this binary may offer initial clarity, it fails to capture the diversity of student engagements now visible in practice. Some students treat AI as a brainstorming partner; others see it as a tool for efficiency, a shortcut to circumvent effort, or even as a disruptive presence to be avoided. Often, these roles shift within the same project. Such variability suggests that AI's role in learning is not fixed but dynamically shaped by epistemic expectations, design constraints, and institutional culture [11, 12].

This paper addresses this conceptual and pedagogical gap by introducing a three-way taxonomy of AI roles (tool, teammate, or neither) implemented across a semester-long design foundation course at the Singapore University of Technology and Design. The course enrolled approximately 500 first-year engineering and architecture students and embedded Gen-AI into the design thinking process, from user research to ideation, prototyping, and testing. Rather than treating AI use as an implicit or purely technical act, the course prompted students to make role decisions explicit at checkpoints. Did the AI function as a single-purpose tool for speed? As a collaborative partner whose contributions required critical negotiation? Or was its use actively suspended to preserve reasoning depth or empirical fidelity?

To this end, the researchers gave student teams the opportunity to apply for a small pool of reimbursable credits to unlock premium Gen-AI services (e.g., ChatGPT Plus, Midjourney, Roboflow). We then required teams to log the use of these tools under three prompts: (1) which platform was used and why, (2) when and how the AI ultimately acted as a tool, a teammate, or was refused, and (3) how that decision affected the design process and outcome (e.g. in terms of time, quality, or ethics). These role decisions were embedded in reflection rubrics, illustrated briefs, and a new Design-AI Award scheme. The goal was to scaffold metacognitive awareness of Gen-AI as both object and subject of design, triggering reflective learning on the different roles Gen-AI can take in students' work [13, 14]. The study triangulates three data sources: (i) thirteen structured reflection spreadsheets of participating teams, capturing tool choice, task allocation, and outcomes; (ii) eight illustrated award briefs distilling teams' key learning moments; and (iii) instructor observation notes from studio facilitation.

By analyzing these materials, we observed how engineering students used Gen-AI not only to increase efficiency but also to cultivate reflexive design judgment, develop hallucination-aware routines, and negotiate task boundaries with a nonhuman collaborator. Through open and axial coding, eight emergent themes were identified: acceleration, collaborative ideation, technical up-skilling, iteration speed-up, known limits, selective non-use, prompt-craft mastery, and AI-embedded artefacts. These themes were mapped onto the proposed taxonomy to assess both its descriptive and analytic utility, and are explored in greater detail in the Findings section.

The study offers two core contributions to engineering education. First, it tests whether a role-based taxonomy of Gen-AI use is observable across authentic student work. Second, it explores how these roles distribute across novice projects in a way that supports the teaching of deliberate, context-sensitive AI integration. More broadly, the study argues that role selection can and should be taught as a form of epistemic design habit. It facilitates a deeper discussion on Gen-AI in education, by moving beyond the question whether students should use such tools towards a more nuanced exploration of how they do so and with what potential benefits to their learning. In the age of Gen-AI, engineering pedagogy must move beyond merely teaching students the technical proficiency how to use these tools. It must also teach them how to decide when to use and when to refuse them.

## II. METHODS

### A. Research Design

This pilot follows a Design-Based Research (DBR) logic that couples instructional innovation with iterative inquiry. The instructional move was the introduction of a three-way AI-role taxonomy (tool, teammate, neither) which was identified in earlier research of the team [e.g. 5, 12, 14]. In addition, a small credit system that let each team purchase premium Gen-AI services was introduced. Data were gathered during the same semester in which the intervention ran, giving real-time insights and allowing for adjustment of reflection prompts where necessary.

### B. Context of study

The setting is *Design Thinking & Innovation* (DTI), a compulsory 13-week module for approximately 500 first-year engineering and architecture students at the Singapore University of Technology and Design. Students work in 4- to 5-person teams across studio classes and lectures, following the UK design council's double-diamond design framework and producing a functional prototype by Week 13. Teams were created by both students and instructors, with a focus on combining engineering and architecture students and on finding team members with complimentary skills. This, to stimulate cross-disciplinary learning and more innovate solutions to addressing the design problem.

The design brief entailed identifying, addressing, and solving an issue on campus as experienced by different users (e.g. students, faculty, staff, cleaning personnel, etc.). One of the learning objectives of the course was to develop proficiency in utilizing AI tools for design tasks and cultivate critical thinking skills regarding the appropriate and effective use of AI in the design process.

The key AI-based interventions made to the course to address these learning objectives were (1) AI based methods were augmented to all design methods, from user research to prototyping. (2) Talks on custom GPTs and AI tools for visualization and prototyping were conducted to encourage students' use of AI. (3) An 'AI x Design thinking' competition was incorporated, to encourage students to reflect on their AI use (4) AI subscriptions were made available, to facilitate equitable access to AI subscriptions.

Each team was given the opportunity to receive S$60 in reimbursable credits to be used for subscriptions to premium Gen-AI platforms (e.g., ChatGPT Plus, Midjourney, Roboflow). Teams that took up this opportunity were asked to log their specific use of the platforms and submit a final reflection at the end of the project. We asked them to elaborate on:

1. Platform(s) used and rationale;
2. When/whether the AI ultimately functioned as tool, teammate, or was refused. The research team provided provisional definitions for these roles in the design brief [see also Figure 1];
3. Perceived effect on time, quality, or ethics.

The competition was conducted by the research team in collaboration with the course coordinators over a span of six weeks. For this competition, teams were asked to reflect on their use of Gen-AI as tool/teammate/neither, providing concrete examples from their design process as illustrations. A competition brief was provided to students with the objectives, AI roles and definitions explained (see Figure 1).

**1 CONTEST OVERVIEW**

This contest challenges you to demonstrate how you used AI creatively and effectively in your project. Your team must showcase how AI was used in the design thinking process. Highlight how these helped you to innovate, and giving specific attention to the three different roles AI can have: tool, teammate, neither.

1. **AI as a Tool** 🔧
   - **Role:** AI assists with tasks by providing data, insights, or suggestions
   - **Focus:** Leverage AI for efficiency and support
2. **AI as a Teammate** 👥
   - **Role:** AI becomes an active participant in creation, collaborating closely with your team.
   - **Focus:** Emphasize synergy between human intuition and AI capabilities.
3. **AI as Neither** 🚫
   - **Role:** AI is either deliberately excluded to emphasize human-driven problem-solving or critically assessed for its limitations.
   - **Focus:** Highlight the value of human-centered design or a balanced critique of AI.

*Figure 1. AI role definitions of tool/teammate/neither used to guide student competition entries*

*C. Data collection and analysis*

This study involved analysis of student coursework artefacts and voluntary reflections. Participation in the reflective-log study and award submissions was strictly voluntary and not linked to grading; non-participation had no academic consequences. We analyzed three artefact sets:

- **Reflection spreadsheets:** (n = 13) log entries covering platform choice, task type, and role tag.
- **Award briefs:** (n = 8) two- to three-page illustrated summaries of AI reflections, submitted voluntarily for jury review.
- **Instructor observations:** the research team checked in with the course coordinators at multiple junctures during the study, providing an additional perspective and, where required, changes to the DBR logic of the study.

Participation in the reflective-log study was self-selected. Teams that wished to unlock S$60 of reimbursable premium Gen-AI credits were required to submit a structured reflection spreadsheet; this yielded *n = 13* team logs. Separately, all students were invited to enter a newly created Design-AI Award. We received *n = 8* illustrated page briefs, of which three were eventual awardees. All eight briefs were included in the qualitative analysis together with the 13 logs and instructor notes. While the sample therefore represents teams highly motivated to engage with Gen-AI, the self-selection aligns with our research focus on *deliberate* AI integration and provides a rich window into high-engagement practices.

The dataset was analyzed using open and axial coding. This structured DBR approach ensured that the taxonomy was both a live pedagogical scaffold and an object of empirical validation. First-cycle codes captured the declared role per AI episode; second-cycle coding grouped rationale statements into thematic families (e.g., acceleration, selective non-use). The next section will detail these different thematic families along the lines of the different roles that students gave to Gen-AI in their designs, illuminating emerging practices.

III. FINDINGS

Drawing from our dataset of student teams reflections and triangulated with insights from project submissions to the competition, we elaborate the three broad roles that AI played in students' design processes: Tool, Teammate, and Neither. These roles do not represent static categories, but rather dynamic stances students took at various phases of their projects. Each was marked by distinctive rationales, practices, and reflections. They reveal not only *what* students did with AI, but *how* they thought about it, when they embraced it, and when they did not. For clarity, in Table 1, we briefly summarize what we mean by each role, what it captures, and provide an illustrative quote.

TABLE 1. ROLES, DESCRIPTIONS, ILLUSTRATIONS

| Role | Description | Illustration |
|---|---|---|
| Tool | AI used for bounded, efficiency-focused tasks (quick concept visuals, parts-list generation, code snippets, sentiment classifiers). | "While emphasizing collaboration, we also utilized AI as an efficient tool for specific tasks." (Team 1) |
| Teammate | AI adopted as a co-designer that debates, critiques and teaches new techniques. The output of AI can be used as input for discussions or further iterations. | "During discussions, we input identical prompts across all models, using the resulting responses to cross-validate ideas, check for hallucinations, and surface blind spots through disagreement." (Team 13) |
| Neither | Intentional suspension or rejection of AI when empathy, context or trust became central. | "Our design aspect of the product was entirely handmade, as we deemed that AI-generated designs are usually very generic and don't give our product a unique identity." (Team 5) |

*A. AI as tool: Delegate drudge, unlock creativity*

The Tool role was used in a number of design tasks, but most evident during visual ideation, code generation, and hardware debugging. When students cast AI as a Tool, they tended to operate with a logic of bounded delegation: assigning repetitive, computational, or technically unfamiliar tasks to AI, thereby freeing up time and cognitive energy for higher-order design work. For example:

*"Using AI tools…transformed both the speed and depth of the design process…This not only saved time but allowed us to experiment with more complex functionalities, such as real-time detection and adaptive motion triggering based on bird presence." [Team 4]*

Rather than automating the design process wholesale, students engaged in deliberate scoping: using AI for what it was good at and containing its reach where precision or empathy was needed.

One team used GPT-4o and HyperSketch to translate rudimentary napkin sketches into high-fidelity visual mockups within hours (see Figure 2). This compressed the traditional sketch-to-CAD cycle and enabled faster prototyping.

*"AI was strategically integrated throughout the industrial design process. First as a tool to generate rapid concept visuals using GPT-4o and HyperSketch, reducing ideation-to-render time from days to under an hour" [Team 14]*

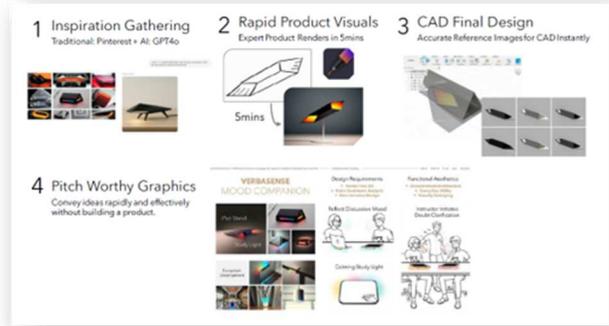

*Figure 2. AI-assisted design workflow: students used GPT-4o and HyperSketch to accelerate visual ideation, generating high-quality renderings and CAD references within hours.*

More importantly, beyond just saving time it allowed teams to explore a greater diversity of ideas and conduct multiple iterations of a prototype, expanding their scope for choosing the most novel design. Teams reported re-investing the time saved into more creative or empathetic aspects of the project. As one team put it:

*"This not only saved time but allowed us to experiment with more complex functionalities, such as real-time detection and adaptive motion triggering." [Team 4]*

Novice coders leaned heavily on AI for troubleshooting. Lacking deep experience in Arduino or C++, they relied on LLMs to spot errors, suggest syntax fixes, and explain electronic concepts in simplified terms. In doing so, AI became a kind of scaffolding helping them cross a knowledge threshold.

*"All of us are new to coding, and debugging is something we are not sure how to do. The AI is able to help us debug successfully." [Team 10]*

Crucially, when engaging AI as a Tool, students also discovered the limitations of the technology in this specific role. Students, for instance, discovered that prompt precision determined the quality of outputs. One team initially used broad, open-ended prompts, but found the results generic and unhelpful. Through trial, feedback, and refinement, they

began treating prompt-writing as its own design task, an emergent form of "meta-craft."

*"Open-ended prompts yielded fairly generic or uninspired results… crafting precise prompts [helped] to guide the AI beyond generic solutions towards identifying analogous operational principles or mechanisms elsewhere." [Team 1]*

These insights strongly echo throughout our dataset. Reflecting on the capabilities and limitations of AI as a Tool, prompt engineering emerged as a key skill with teams learning to refine inputs iteratively or even using one AI model to generate prompts for another. In short, when using AI as a Tool, students used AI to amplify their productivity, but not at the cost of critical oversight.

### B. AI as teammate: Conversational partner and co-designer

In contrast, the Teammate role involved a qualitatively different relationship: AI was treated not as an assistant to perform in a specific task, but as a dialogical partner. In some instances, it even became an additional team member capable of critique, debate, and tutoring. This new 'team member' could address skill or expertise gaps missing in the team. For instance, several teams described using AI to overcome unfamiliarity with electronics and programming, crediting it with helping them debug code successfully and accelerate hardware development.

One of the more compelling examples of AI as a critical teammate came from a team that ran the same design prompt through four different LLMs (Gemini, ChatGPT, DeepSeek, and Claude) and then conducted a collective review of the responses.

*"We input identical prompts across all models, using the resulting responses to cross-validate ideas, check for hallucinations, and surface blind spots through disagreement." [Team 13]*

This practice of "hallucination fire-drills" transformed AI from a black box tool into a collaborative teammate. Disagreement among models became a site for metacognitive inquiry. It helped students challenge their own assumptions, identify blind spots, and even explore novel solutions.

What distinguishes the Teammate role is the reflective layering of AI interactions that it triggers. Rather than using AI for single-output queries, students began constructing comparative conversations across models or recursive critiques within a design phase. One group [Team 1], described how their "most significant use of AI positioned it as a collaborative partner in the design process." They fed initial ideas into an LLM to "analyze the inherent strengths and weaknesses of each proposed mechanism," and used those critiques to debate and reframe their physical prototype. This process was repeated multiple times, with students iteratively adjusting prompts and cross-referencing suggestions with teammates' insights, prompting rapid, structured critique, "much like an additional team member offering critical perspectives."

In curating dialogue between different models, and between themselves and the models, students moved from being passive AI users to active AI orchestrators, deliberately steering machine conversations to deepen their design reasoning. These recursive cycles mirror metacognitive scaffolding, where learners surface their reasoning, anticipate flaws, and evaluate alternative interpretations.

In some instances, AI moved from ideation partner to embedded system, with different AI platforms being linked up. Team 5, for example, built a smart assistant designed to enhance classroom engagement. In this project, AI was not simply used to generate content or automate tasks, but it was woven into the design process as a collaborative partner across multiple stages. Their system integrated OpenAI's Whisper-1 model for real-time classroom transcription, GPT-4 for lecture summarization, and a context-sensitive GPT chatbot that enabled students to ask follow-up questions. The chatbot's answers were dynamically grounded in the day's transcript and summary, effectively turning it into a responsive classroom aide. Rather than relying on AI merely for back-end processing, the team described how "AI shaped every stage of our process" and "served as not just a tool but a creative collaborator." Their approach demonstrates technical layering (speech-to-text, summarization, and conversational agents) but also added a layer of support for understanding, as AI was used to explain, interpret, and connect classroom content for students.

Together, these examples show that when students positioned AI as a teammate rather than purely a task tool, they explored new forms of collaborative learning. Here, critique, curiosity, and cross-checking were central to the design process.

### C. AI as neither: Knowing when to say no

Perhaps the most striking stance taken by students was the deliberate choice to refuse AI at key junctures. This "Neither" role was not a rejection of AI's value, but a situated decision. Students recognized that some tasks demanded *presence*, *empathy*, *ethical judgments* or *trust* that AI could not yet replicate, or that students wanted to keep under their own influence. For instance, a team developing an autonomous dining tour car, reflected:

*"We conducted direct observational analysis, student surveys, and interactive campus walkthroughs—invaluable qualitative approaches AI struggled to replicate fully. For instance, selecting scenic pauses at locations such as the Jackie Chan Pavilion [a monument on campus] required human intuition around cultural appreciation and emotional*

*resonance—nuances difficult for an AI to perceive."*
*[Team 3]*

Teams frequently paused AI interaction during user interviews, stakeholder conversations, and participatory workshops. In these "empathy zones," which are crucial for the process of user journey mapping or user experience in design projects, students viewed AI as potentially intrusive or reductive. Team 14 used their AI platforms in all its three roles in understanding user personas. Reflecting on the process, they mention AI was used as a tool for rapidly identifying needs and problems in their collected corpus of data, while it acted as a teammate during actual user interviews by applying real-time sentiment analysis, offering an additional layer of depth. However, in their conduct of the actual empathy mapping and when doing follow-up interviews, "AI was deliberately excluded…to preserve human nuance and ensure unbiased, authentic insights."

Interestingly, while AI is often seen as an impediment to learning (e.g. because students can easily "offload" their learning to such a technology), we also saw that some students were themselves aware of it and thus decided to not use AI to retain their learning opportunities. For some, this was in the context of user research, and for others this was in the more technical aspects of their projects such as coding or wiring. Team 5, for instance, while admitting that their product relied heavily on technical AI capabilities, wanted to remain being the central actor in terms of the design and aesthetics of it:

*"However, this [use of AI platforms] does not mean that it has entirely replaced our roles in this project. Our design aspect of the product was entirely handmade, as we deemed that AI generated designs are usually very generic and don't give our product a unique identity. In this regard, we manually came up with our software & web interface design and 3D model to print." [Team 5]*

Finally, at times AI's role as Neither was reinforced by moments of misfire. One team asked ChatGPT and DeepSeek for a list of electronic parts to build their prototype: a portable projector that users could interact with. Trusting the AI, they bought everything it suggested.

*"Significant mistakes were made following the advice AI had given. We were painfully made aware of the stakes of relying on AI too much, leading to us spending much more money than initially expected." [Team 15]*

Several components turned out to be incompatible or difficult to use. For example, the projector module didn't work properly with their setup, even though ChatGPT had recommended it. Sensors and chips it suggested also proved too complex to integrate. In the students' words, the AI made the project sound easier than it really was. Instead of giving up on AI, the team became more cautious. They began cross-checking AI suggestions with expert advice and real-world testing before making decisions. Over time, they learned to treat AI not as an authority, but as one tool among many.

## IV. DISCUSSION AND IMPLICATIONS

This section returns to the central question posed in the title: to use or to refuse? This phrase encapsulates a key pedagogical strategy, that prompts students to make deliberate, context-aware decisions about when to engage generative AI, how to use it, and when to set it aside. Rather than accepting AI as a default solution, students were taught to pause, assess, and justify their choices. This structure directly addresses the common critique that AI may lead to cognitive passivity or superficial learning. Instead, our findings show that when role selection is embedded in design pedagogy, AI can serve as a catalyst for deeper reflection, ethical consideration, and iterative creativity. The section that follows unpacks what this decision-making process means for theory, practice, and future inquiry in engineering design education.

### A. Role selection as a design pedagogy

The findings reveal that when students are explicitly asked to reflect and categorize AI interactions as tool, teammate, or neither, they develop more sophisticated decision-making frameworks that extend beyond the automation-versus-augmentation binary identified earlier in the literature [8, 12]. While this binary framework provides initial clarity about AI's potential roles, our study demonstrates that it inadequately captures the dynamic, context-sensitive nature of student-AI interactions observed in authentic design contexts.

The tool-teammate-neither taxonomy addresses the conceptual gap identified in the introduction: that AI's role in learning is not fixed but dynamically shaped by epistemic expectations, design constraints, and institutional culture [11]. Our findings show that students naturally shifted between these roles within the same project, suggesting that role fluidity rather than role consistency characterizes effective AI integration in design education.

Importantly, the "neither" category challenges the assumption that AI adoption should always be maximized. Instead, it reveals that deliberate non-use represents a sophisticated form of technological literacy. Students who chose to refuse AI during specific parts of their work, such as empathy mapping or user interviews, demonstrated awareness of the technology's limitations and preserved the authenticity of human-centered design processes. This finding directly responds to concerns about "skill erosion" raised in institutional responses to ChatGPT [6, 7], showing that students can maintain critical judgment when given appropriate scaffolding.

We therefore hypothesize that deliberately asking students to consider and reflect on the role that Gen-AI takes in their project surfaces metacognitive moves. These would arguably remain tacit under traditional instruction, where the use of AI would either be a matter of developing the required technical skills or where the role would have been mandated by the instructor.

### B. Beyond prompt engineering: Metacognitive AI literacy

The study reveals that focusing solely on prompt crafting, which is a common emphasis in early AI education initiatives, misses the deeper metacognitive skills required for effective AI integration. Students who treated AI as a teammate developed what we term "hallucination-aware workflows," using disagreement between models as a site for critical inquiry rather than viewing AI outputs as authoritative.

This builds on calls for moving beyond technical proficiency toward teaching students how to decide when and how to use AI tools, and when to turn them off. The metacognitive awareness demonstrated by students who conducted "hallucination fire-drills" represents a design habit that addresses concerns about AI leading to cognitive passivity or superficial learning. By developing these sophisticated decision-making frameworks, students transformed from passive AI user into active AI orchestrators, demonstrating that a crucial aspect of AI literacy involves knowing not just how to use these tools, but when to critically evaluate, strategically limit, or deliberately refuse them.

### C. Implications for engineering education and institutional policy

The findings have direct implications for institutional responses to AI in education. Rather than the "ambivalent" approaches documented in early literature, characterized by concerns over academic integrity, skill erosion, and the difficulty of assessing AI-mediated work [6, 7], our study suggests that structured integration with explicit role reflection can transform AI from a threat to academic integrity into a catalyst for deeper learning and design innovation.

The coordinated approach we implemented (combining tool access, reflection requirements, role tagging, and public recognition through the design competition) demonstrates how institutions can move beyond "compliance and containment" toward productive AI integration. This addresses the limitation identified in early discourse that focused primarily on whether students use Gen-AI and a more normative stance on whether students should use it rather than exploring how students position AI within their learning practice.

The study demonstrates that AI role selection can be made "assessable" through structured reflection rubrics and illustrated briefs. This addresses the challenge of assessing AI-mediated work identified in early institutional responses [6, 7]. By requiring students to justify their role choices and reflect on outcomes, educators can evaluate not just the final product but the decision-making process itself. The self-selected nature of our sample (teams highly motivated to engage with AI) suggests that scaling this approach institution-wide would require additional scaffolding for less motivated students. However, the rich practices observed among engaged students provide a foundation for developing more comprehensive AI literacy curricula.

## V. LIMITATIONS AND FUTURE RESEARCH

This pilot study has several limitations that suggest directions for future research. First, the self-selected sample represents highly motivated students, potentially limiting generalizability. Future studies should examine how the tool-teammate-neither framework applies across different levels of student engagement and motivation. The framework itself should also be validated with additional experts and research. While the competition brief provided definitions of the different roles of AI, a more formal classification of some usage is still open to interpretation. Because much of our evidence consists of self-reported reflections, responses may be shaped by social desirability or perceived instructor expectations. We mitigated this by decoupling the submissions from students' grades, but even so, self-report bias cannot be ruled out.

Second, the study focused on a single engineering design course. Research is needed to understand how role-based AI integration transfers to other engineering disciplines and educational contexts. Third, the semester-long timeframe provides a snapshot of student learning but cannot capture longer-term impacts on professional development or career preparation.

Future research should also investigate how the taxonomy performs across different AI technologies as they evolve, and examine whether the role-based approach maintains its effectiveness as students become more familiar with AI tools. Additionally, studies should explore how instructor preparation and institutional culture influence the success of structured AI integration approaches.

## VI. CONCLUSIONS

This study demonstrates that the core question facing engineering education should not be *whether* to allow AI use among students, but *how* to teach students to use it thoughtfully. By implementing a tool-teammate-neither taxonomy within a design thinking curriculum, we observed that students can develop sophisticated AI literacy that goes beyond technical proficiency to include metacognitive awareness, ethical reflection, and strategic decision-making.

The findings challenge the automation-versus-augmentation binary that dominated early AI education discourse, revealing instead that effective AI integration requires dynamic role selection based on task requirements, ethical considerations, and learning objectives. Students who learned to use AI as a tool, as a teammate, and to deliberately refuse AI at key junctures, demonstrated a high level of technological literacy, suggesting that selective non-use should be valued equally with strategic use.

As generative AI continues to reshape engineering practice, education must evolve beyond teaching students how to use these tools to teaching how to think about using them. The approach explored here offers one pathway toward this goal. By re-centering student agency in AI integration decisions, we can harness the technology's potential while preserving the critical thinking and human-centered design skills that remain essential to engineering practice.

The implications extend beyond individual classrooms to institutional policy and curriculum design. Rather than viewing AI as a threat to be contained, educational institutions can embrace it as an opportunity to develop new forms of literacy and reflection. In doing so, they can prepare students not just for a world with AI, but for a world in which knowing when to use AI *and* when to refuse it becomes a fundamental professional skill.

## ACKNOWLEDGMENT

This research is supported by the Ministry of Education, Singapore, under its Tertiary Education Research Fund (MOE2024-TRF-30). Any opinions, findings and conclusions or recommendations expressed in this material are those of the author(s) and do not reflect the views of the Ministry of Education, Singapore.